\documentclass[11pt,a4paper]{article}
\usepackage[left=20mm, top=15mm, right=20mm, bottom=25mm, nohead=true, nofoot=true]{geometry}
\usepackage[utf8]{inputenc}
\usepackage{authblk}
\usepackage{indentfirst}
\usepackage{misccorr}
\usepackage{graphicx}
\usepackage{amsmath}
\usepackage{amssymb}
\usepackage{multicol}
\usepackage{bm}
\usepackage{physics}
\usepackage{longtable}
\usepackage{pdflscape}
\usepackage{epstopdf}
\usepackage{multirow}
\usepackage{adjustbox}
\usepackage{amsfonts}
\usepackage{ifthen}
\usepackage{fancyhdr}
\usepackage{setspace}
\usepackage{xcolor}
\usepackage[round,authoryear,comma,sort&compress,numbers]{natbib}
\usepackage[toc,title,page]{appendix}
\usepackage[hidelinks]{hyperref}
\usepackage{url}

\hypersetup{
    colorlinks=true,
    linkcolor=green,
    citecolor=blue,
    filecolor=magenta,
    urlcolor=cyan,
    pdftitle={Sharelatex Example},
    pdfpagemode=FullScreen,
}

\fancypagestyle{plain}{\chead{\texttt{\textcolor{brown}{Communications of BAO, Vol.68, Issue 1, 2021, pp.121-124}}}}
\title{\textbf{Coupling and recoupling of binaries in chaotic three body systems}}
\author{T.S.Sachin Venkatesh\thanks{tssachin.venkatesh@gmail.com}}
\affil{\scriptsize Delhi Technological University, New Delhi, India}

\def\apj{Astrophys.~J. }

\def\nat{Nature. }

\begin{document}
\pagestyle{empty}
\newpage
\pagestyle{fancy}
\label{firstpage}
\date{}
\maketitle

\begin{abstract}
Three body systems where one of the bodies is ejected without escaping the binary system have previously been studied in various restricted forms. However, none of these studies dwells on the problem in a general setting. Thus, to study this phenomenon qualitatively, we try to expand this problem's scope to unequal mass systems and generalize them by considering various configurations of fixed initial points with precisely calculated initial velocities, some zero velocity models and some optimized models. We will see the use of  terminology similar to the previous studies done in this domain but incorporate different analytical and evaluation methods.
\end{abstract}

\section{Introduction}
The three-body problem is one of the long-standing fundamental problems of dynamical systems. Until 1975, there were not many models of three-body systems and the ones that existed were too restrictive and specific about the parameters of the model. In 1890, Poincaré proved the non-existence of the uniform first integral of a three-body problem in general, and also highlighted the sensitive dependence to initial conditions of its trajectories, a brief account of the same is given in \cite{poinc}. Three-body systems without mass hierarchy are never thought to be stable for very long. They can indeed exist for some time, but they are not found to be stable long term. In these systems, each body orbits the center of mass of the system. Mostly, two of the bodies form a close binary system, and the third body orbits this binary at a distance much larger than that of their orbit. This arrangement is called hierarchical. The reason for this behavior is that if the inner and outer orbits are comparable in size, the system may become dynamically unstable, leading to a body being ejected from the system. \citet{hutbahcall} had done a qualitative study on such ejecting systems. However, their model was restricted to an equal mass system in 2 dimensions specifically modeled according to resonance scattering. They also stated that generalizing that specific problem to 3 dimensions would not provide new insights as they would qualitatively be the same. 

\section{Methods}
We focus this work on chaotic systems where the three body system finally reduces to a binary and an \textit{escaper}. We then compare them to a benchmark, a stable unequal mass three body system on the basis of their phase space, the total energy of the system and the evolution of different binaries in the systems. The stable system used as a benchmark has a mass ratio of 1:1:2 and a cyclic phase space demonstrating that the system is stable over a large number of iterations. Due to its computationally expensive nature, the system was simulated for only 500 iterations. The test systems simulated were constructed using the dataset provided by \citet{data} for different configurations of Plummer models and an extensive list of initial positions. All simulated systems initially exist in a temporarily bound triple mass system.

We notice the use of Euler-Cromer method to account for the accumulation of errors in the total energy of the model, keeping in mind the instability in the system. To minimize error propagation in position and velocity space, the Levenberg-Marquardt method was used to account for minuscule and extreme values as artificial values creep in over longer iterations in the previous method which cannot be tolerated in the case of these values.

With the kinetic energy of the system, $E_{kin}$ given by
$$E_{kin} =\frac{1}{2}\sum_{i=1}^{3} M_{i}\dv{R_{i}}{t}.\dv{R_{i}}{t}$$
and the potential energy, $E_{pot}$ given by
$$E_{pot} = -\frac{G}{2}\sum_{i=1}^{3}\sum_{j=1}^{3}\frac{M_{i}M_{j}}{r_{ij}},$$
we have total energy $E_{tot} = E_{kin}  + E_{pot}$

The simulations were stopped when the systems dissolved into a binary and an \textit{escaper}. We will categorize a body as an \textit{escaper} along the same lines as \citet{data}. The conditions are :
\begin{itemize}
  \item The \textit{escaper} has a positive energy, \textbf{$E_{esc}$ $>$ 0}
  \item The \textit{escaper} is moving away from the centre of mass, \textbf{r.v $>$ 0}
\end{itemize}

The barycentric frame of the bodies were considered when calculating the energy of the \textit{escaper}. We do not differentiate between hard binaries and soft binaries as it has been stated by \citet{binaryproof} that these metrics do not do justice and do not provide enough information in the case of unequal mass triple systems. An upper limit of 500 iterations is set, to not exceed the benchmark and a lower limit of 200 iterations. Models that dissolved into a binary and an \textit{escaper} before 200 iterations were not included in the sample space.

\begin{figure}[htp]
    \centering
    \begin{minipage}{0.45\textwidth}
        \centering
        \includegraphics[width=0.9\textwidth]{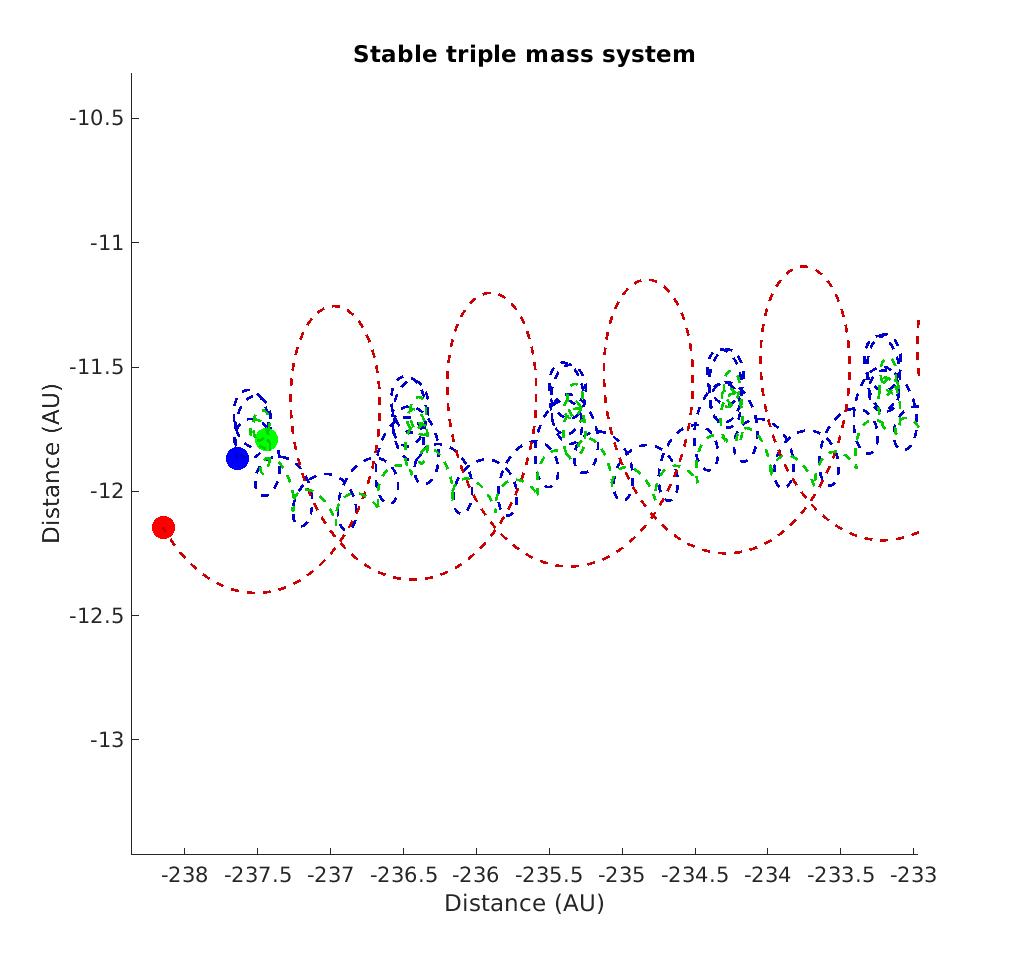} 
        {Stable three body system, the benchmark for this study. The mass ratio of the bodies are 1:1:2 (Blue:Red:Green).}%expand this?
    \end{minipage}\hfill\\
    \vspace{10pt}
    \begin{minipage}{0.45\textwidth}
        \centering
        \includegraphics[width=0.9\textwidth]{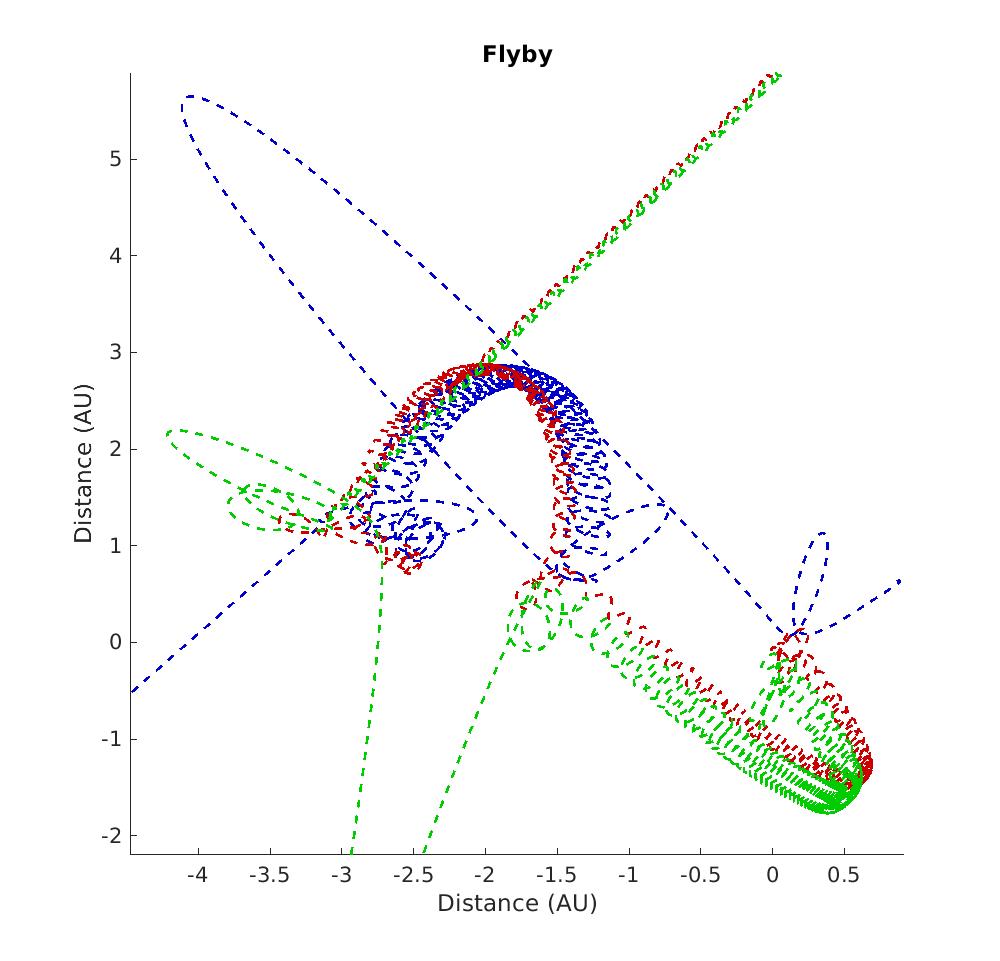} 
        {\\The initial binary is formed by the red and green bodies while the blue body traverses a returning trajectory. Splitting and recoupling continue until the system permanently splits into the initial binary and an \textit{escaper}.}
    \end{minipage}
    \begin{minipage}{0.45\textwidth}
        \centering
        \includegraphics[width=0.9\textwidth]{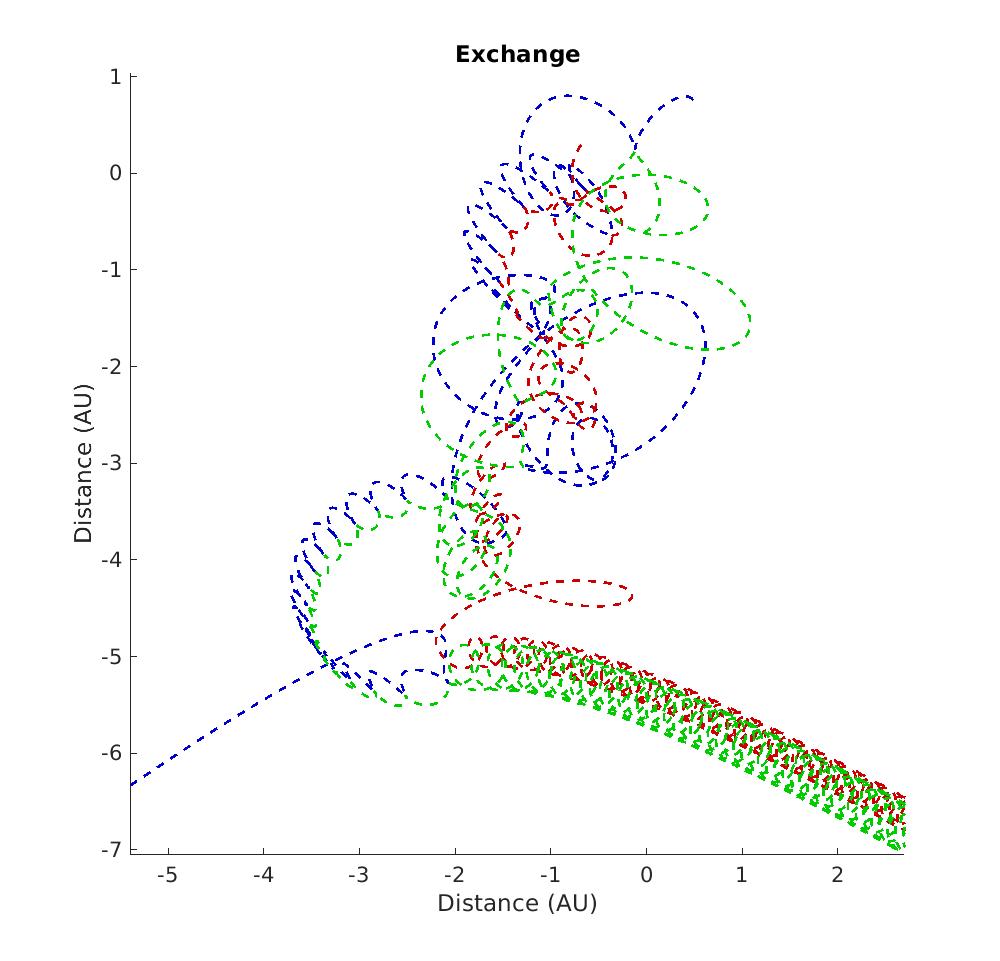} 
        {\\The initial binary is formed by the red and blue bodies, which is cut short by the body in green. Splitting and recoupling continue until the system permanently splits into the non initial binary and an \textit{escaper}.}
    \end{minipage}\
    \caption{The code written by \citet{sim} forms the basis for all the simulations. We added total energy calculation for both the system and the bodies and a phase diagram grapher to this program. This helps predict the successive motion of the bodies and thus, the stability of the systems designed using data from \citet{data}.}
\end{figure}

\section{Observations and discussion}
Of all the systems simulated, some displayed a peculiar behaviour where a body is ejected without escaping from the binary. After a long solitary path, such bodies turn around and engage the binary system once again to form a temporarily bound triple system. This mechanism will be called as \textit{quasi-ejection} henceforth. Based on the evolution of the system, if it dissolves into a binary and an \textit{escaper} after being in a temporarily bound triple system for $iterations > lower\hspace{3pt}limit$, we categorise them accordingly :
 \begin{itemize}
     \item If the end binary of a system is the same as the binary formed after the first \textit{quasi-ejection}, we call it a \textbf{\textit{flyby}}
     \item If the end binary of a system is not the same as the binary formed after the first \textit{quasi-ejection}, we call it an \textbf{\textit{exchange}}
 \end{itemize}

The ratio of \textit{flybys} to \textit{exchanges} was practically equal across all system configurations, with some bias towards \textit{flybys}. However, the result of importance here is that the \textit{escaper} turned out to be the lowest mass body in most cases and the end binary constituted the more massive bodies. If $E_{tot} > 0$, the system must split and if $E_{tot} < 0$, the system may lead to an \textit{escaper} or form a stable three body system. But the combined constraints of $E_{tot}$ of the system and the dynamic exchange of $E_{tot}$ between each body point towards the lowest mass body being the \textit{escaper} in most cases. This is complimented by the result deduced from $E_{esc}$.

We suspect the mass hierarchy of the bodies to be the primary governing factor in this problem of coupling and recoupling of binaries, not the initial conditions as one may believe. In \textit{Exchanges} where the initial binary constituted of the lowest and highest mass bodies and the \textit{escaper} was the highest mass body were extremely low, especially in disproportionate mass systems. This can be attributed to the Hill sphere \citep{hillsp} mechanism, thus destabilizing the binary. Nevertheless, in almost all cases, the new binary is formed by the massive bodies which attain stability, and the lowest mass body ends up being the \textit{escaper}. Given the wide scope for further studies in this sub-domain and the large impact it can have on qualitative analysis of similar systems in the future, and therefore is likely to increase our understanding of the mechanisms and methods at play, preliminary results have been presented here. To go beyond these results requires a more detailed analysis and robust computations, which are postponed to a future article.

\section*{\small Acknowledgements}
\scriptsize{Special thanks to Suyog Garg and Shobhit Ranjan for their insights on the article and for providing useful comments.}

\scriptsize
\bibliographystyle{ComBAO}
\nocite{*}
\bibliography{references}

\end{document}